\documentstyle[twoside,fleqn,espcrc2]{article}

\title{ \vspace{-3.2cm}
\begin{flushright}
{\normalsize IFUP--TH 46/96} \\
{\normalsize July 1996} \\
\end{flushright}
\vspace{1.5cm}
Functional integration on Regge geometries}
\author{Pietro Menotti and Pier Paolo Peirano \\ \mbox{}
\\ Dipartimento di Fisica dell'Universit\`a, 56100 Pisa, Italy and
INFN Sezione di Pisa}

\begin{document}

\begin{abstract}
We adopt the standard definition of diffeomorphism for Regge gravity
in $D=2$ and give an exact expression of the Liouville action in the           
discretized case. We also give the exact form of the integration
measure for the conformal factor. In $D>2$ we extend the approach to
any family of geometries described by a finite number of
parameters. The ensuing measure is a geometric invariant and it is
also invariant in form under an arbitrary  change of parameters.
\end{abstract}
\maketitle

\section{Introduction}

In the following we shall consider the Regge manifold as a
differential manifold equipped with a singular metric which is
everywhere flat except on $D-2$ dimensional simplices. A singular metric
does not conflict with the differential structure.  In fact the
concept of differential manifold precedes that of Riemannian manifold
\cite{koba} which means that the charts and transition functions are to be
given before equipping the manifold with a metric; once the
differential manifold is given one can consider also singular metrics
on it \cite{chee}.

In defining the functional integral the choice of the local
fundamental variables plays a key role. The analog of the connection
$A_\mu$ of Yang-Mills theory is played in gravity by the metric tensor
$g_{\mu\nu}$ and the gauge transformations are replaced by the
diffeomorphisms. Similarly to what happens in the finite dimensional
case, the integration measure is induced by a distance in the space of
the field configurations. Such a distance must be invariant under the
relevant symmetry group of the theory and ultralocal; this last
property is implied by the request that the integration measure
should play a kinematical and not a dynamical role. In the case of
gravity the most general distance which is invariant under
diffeomorphisms and ultralocal has been given by De Witt \cite{dewitt} 
\begin{equation}
\label{dewittm}
(\delta g,\delta g) = \int \sqrt{g}\, d^D x \,\delta
g_{\mu\nu}
G^{\mu\nu\mu'\nu'} \delta g_{\mu'\nu'}
\end{equation}
with
\begin{eqnarray}
G^{\mu\nu\mu'\nu'} &=& g^{\mu\mu'}g^{\nu\nu'} +
g^{\mu\nu'}g^{\nu\mu'}  \\ &  - &
{2\over D} g^{\mu\nu}g^{\mu'\nu'}+C g^{\mu\nu}g^{\mu'\nu'}, \quad C > 0 .
\nonumber
\end{eqnarray}

The main job is that to extract from the integration measure induced
by (1) the infinite volume of the diffeomorphisms. We shall examine
first the simpler case of $D=2$ and then go over to $D>2$.

\section{$D=2$}

We recall that the De Witt metric (1) is the starting point of the
treatment of 2 dimensional quantum gravity on the continuum
\cite{allc} and this should be kept in mind when comparing the results
of the continuum theory with the outcome of numerical simulations. On
the continuum a complete reduction of the functional integral has been
achieved in the conformal gauge \cite{allc} giving rise to the well known
Liouville action
\begin{eqnarray}
\lefteqn{S_{L}[\sigma,\hat{g}(\tau_{i})] =} \label{liouv} \\
& =  \frac{26}{24\pi} \int d^2 x \,
\sqrt{\hat{g}} \; [  \hat{g}^{\mu\nu} \partial_{\mu} \sigma
\partial_{\nu} \sigma + R_{\hat{g}} \sigma ] \nonumber
\end{eqnarray} 
which is the log of the determinant of the Lichnerowicz operator
$P^\dagger P$ computed on the surface described by the conformal
factor $e^{2\sigma}$. 

The Regge scheme where the curvature is
localized at isolated points represents a natural discretization of
the functional integral; it is the analog of restricting in ordinary
quantum mechanics the functional integral, to piecewise linear paths,
and let at the end the number of segments go to infinity. The
conformal factor describing a Regge geometry with the topology of the
sphere is given by \cite{foeraur,pmppp}
\begin{equation}
e^{2\sigma} = e^{2\lambda_{0}} \prod_{i=1}^{N} | \omega - \omega_{i}
|^{2(\alpha_{i} -1)} 
\label{fattore}
\end{equation}
with $0<\alpha_{i}$ and $\sum_{i=1}^{N} ( 1 - \alpha_{i} ) = 2$.
As it happens on the continuum such a conformal factor is unique
\cite{ag} up to the 6 parameter $SL(2C)$ transformations
\begin{eqnarray}
\label{conftras}
& \lambda_{0}' = \lambda_{0} + \sum_{i=1}^{N} (\alpha_{i} -1 ) \log
|\omega_{i} c + d| & \nonumber \\
& \displaystyle \omega_{i}' = \frac{a\omega_{i} + b}{c\omega_{i} + d }
\; ,
\qquad \alpha_{i}' = \alpha_{i}  &
\end{eqnarray}
with the complex parameters satisfying $ad -bc = 1$. Such a description
is equivalent to usual one in terms of link lengths. In
fact the number of parameters in (\ref{fattore}) are $3N$ from which
we have to subtract the dimension of the invariance group obtaining
$3N-6$. This equals the number of links $H$ given by the Euler
relation $H=F+N-2$ with $N$ the number of vertices and $F=2H/3$ the
number of faces. On the other hand it is mathematically more
advantageous because the links represent arcs of a very special set of
geodesics connecting the vertices, among which one has to impose a
large number of triangular inequalities, while the parameters
$\omega_i$ vary without constraints. It is possible to give an exact
expression of the determinant of the 
Lichnerowicz operator on a Regge surface to obtain for the action 
\cite{pmppp} 
\begin{eqnarray}
\lefteqn{S_{L} = \frac{26}{12} \left\{ \sum_{i,j\neq i}
\frac{(1-\alpha_{i})(1-\alpha_{j})}{\alpha_{i}}  \log
|w_{i} - w_{j}| + \right.} & &\nonumber \\
& & \left. + \lambda_{0} \sum_{i} (\alpha_{i} - \frac{1}{\alpha_{i}}) 
- \sum_{i} F(\alpha_{i}) \right\}  
\label{azione}
\end{eqnarray}
where $F(\alpha)$ is given by an integral representation.
Action (\ref{azione}) is invariant under the $SL(2C)$ group and in the
continuum limit goes over to eq.(\ref{liouv}). The discrete
counterpart of the functional measure ${\cal D}[\sigma]$ is given by
\begin{equation}
  {\cal D}[\sigma] = \prod_{k=1}^{N} d^{2}\omega_{k} \;
  \prod_{i=1}^{N-1} d\alpha_{i} d\lambda_{0} \, \sqrt{\det J}
\end{equation}
where $J$ is the determinant of the $3N\times 3N$ matrix 
\begin{equation}
  \label{jac}
  J_{ij} = \int d^{2}\omega \, e^{2\sigma} \, \frac{\partial
    \sigma}{\partial p_{i} } \frac{\partial \sigma}{\partial p_{j}},
\end{equation}
being $p_i$ the parameters $\omega_{1x}\dots \omega_{Ny},$ $\lambda_0,$
$\alpha_1\dots \alpha_{N-1}$. Also such integration measure is
invariant under $SL(2C)$. Both the action and the measure can be
written explicitly also for the torus topology \cite{pmppp}. In this
case the product of the exponential of the action and the measure is
invariant under translations and under the modular transformations. 

Action (\ref{azione}) is not local but very simple; less simple, even 
though explicitly known, is the determinant of $J_{ij}$.

On the numerical front, accurate simulations have been given of two
dimensional gravity, both pure and coupled with Ising spins by
adopting the measure $\prod_i \frac{dl_i}{l_i}$. The results
are consistent with the Onsager exponents and in definite
disagreement with the KPZ exponents \cite{janke} while the situation
for the string susceptibility is still unclear \cite{holmj}. That
measures of type $\prod_i dl_{i} f(l_{i})$ fail to reproduce the
Liouville action can be understood by the following argument \cite{pmppp}:
on the continuum for geometries which deviate slightly from the flat
space one can compute approximately the Liouville action by means of a
one loop calculation. If one tries to repeat a similar calculation for
the Regge model with the measure $\prod_i dl_{i} f(l_{i})$ one
realizes that being the Einstein action in two dimensions a
constant, the only dynamical content of the theory is played by the
triangular inequalities. But at the perturbative level triangular
inequalities do not play any role and thus one is left with a
factorized product of independent differentials which bear no dynamics
and thus no Liouville action. The problem is that the Liouville action
is not the result of integrating on fluctuations of the geometry but
of integrating on the diffeomorphisms, while keeping the geometry
exactly fixed.

\section{$D>2$}

While in $D=2$ the
geometry is completely described by a finite number of parameters plus
a conformal factor, in $D>2$ this is no longer true. We consider a
general scheme in which the $D$--dimensional geometries are described
by a class of metrics $\bar g_{\mu\nu}(x,l)$ depending on a finite
number of parameters $l_{i}$. In the Regge model one can think of the
$l_{i}$ as the link lengths, but any other parameterization or class of
geometries is equally good.
We want to treat exactly the diffeomorphisms $f$ \cite{jev} and thus the
functional integral will be performed on the class of metrics
$g_{\mu\nu} (x,l,f) = f^{\star} \bar g_{\mu\nu}(x,l)$ and again the job will
be that to extract from the  De Witt measure the infinite volume of
the diffeomorphisms. 
To this purpose the variation of the metric will be decomposed in two
orthogonal parts
\begin{equation}
\delta g_{\mu\nu} = (F\xi)_{\mu\nu} + ( 1 - F(F^{\dag}F)^{-1}F^{\dag})
\frac{\partial g_{\mu\nu}}{\partial l_{i}} \delta l_{i} 
\end{equation}
being $(F\xi)_{\mu\nu} = \nabla_{\mu} \xi_{\nu} + \nabla_{\nu}
\xi_{\mu}$  the action of an infinitesimal diffeomorphism.
The adjoint of $F$ according to metric (1) is given by 
\begin{equation}
(F^{\dag}h)_{\nu} = -4 \nabla^{\mu} h_{\mu\nu} - 2 ( C - \frac{2}{D} )
\nabla_{\nu} h^{\mu}_{\mu}
\end{equation}
and the inverse of $F^{\dag} F$ is well defined from Im$(F^{\dag})$
onto itself.
Factoring the volume of the diffeomorphisms one reaches for the
integration measure the expression \cite{gen}
\begin{equation}
\prod_{k} dl_{k} \left( \det(t^{i}, t^{j}) {\cal D} \mbox{et } (F^{\dag} F)
\right)^{\frac{1}{2}} 
\label{misura1}
\end{equation} 
where
\begin{equation}
t^{i}_{\mu\nu} = \left( 1 - F(F^{\dag} F)^{-1}F^{\dag} \right)
\frac{\partial g_{\mu\nu}}{\partial l_{i}} \; .
\end{equation}

The main feature of eq.(\ref{misura1}) is to be a geometric invariant
i.e.\ it does not change under diffeomorphisms which may also depend
on the parameters $l$ and thus under a larger class of transformations
than the De Witt metric. On the other hand it depends through
$F^{\dag}$ on the parameter $C$. 

One expects the dependence on $C$ to disappear when the number of
parameters $l$ becomes large (continuum limit). In fact if one
enlarges the scheme by integrating on a class of metrics $g_{\mu\nu} =
f^{\star} e^{2\sigma} \hat g_{\mu\nu}(x, \tau)$, where the finite
number of parameters $\tau$ describe deformations transverse both to
diffeomorphism and to the Weyl group, one reaches the $C$--independent
measure \cite{gen}
\begin{eqnarray}
\lefteqn{ \prod_{k} d\tau_{k} {\cal D}[\sigma] \left[  {\cal D}\mbox{et} (
  P^{\dag} P ) \cdot  \right. }  & & \label{misura2} \\
  & & \left. \cdot \det \left(  k^{i}, ( 1 - P
  (P^{\dag} P)^{-1} P^{\dag})
  k^{j} \right) \right]^{\frac{1}{2}}
\nonumber
\end{eqnarray}
where $(P\xi)_{\mu\nu} = \nabla_{\mu} \xi_{\nu} + \nabla_{\nu}
\xi_{\mu} - \frac{2}{D} g_{\mu\nu} \nabla \cdot \xi$
and
\begin{equation}
k^{i}_{\mu\nu} = \frac{\partial g_{\mu\nu}}{\partial \tau_{i}} -
\frac{g_{\mu\nu}}{D} g^{\alpha\beta} \frac{\partial g_{\alpha
      \beta}}{\partial \tau_{i}} \; .
\end{equation}
The determinants appearing in eqs.(\ref{misura1}, \ref{misura2})
are both well defined, being $F^{\dag}F$ and $P^{\dag}P$ elliptic
operators, but contrary to what happens in the 2 dimensional case the
dependence on $\sigma$ of 
${\cal D}\mbox{et }(P^{\dag}P)$ cannot be reduced to the computation
of a local variation of $\sigma$, due to the non ellipticity of $P
P^{\dag}$.

In conclusion in $D=2$ an explicit form for the discretized path
integral has been given, satisfying the correct invariance
properties. The scheme can be extended to $D>2$, but the explicit
calculation of the determinants is not straightforward.


\begin{thebibliography}{9}

\bibitem{koba}  S.\ Kobayashi, K.\ Nomizu, {\it Foundations of differential
                geometry}, Interscience Publishers (1963).

\bibitem{chee}  J.\ Cheeger, {\it J.\ Diff.\ Geom.} {\bf 18} (1983) 575.

\bibitem{dewitt}  B.\ S.\ De Witt, {\it Phys.\ Rev.} {\bf 160}, 1113.

\bibitem{allc}  A.M.\ Polyakov, {\it Phys.\ Lett.} {\bf 103B} (1984)
                207; 
                J.\ Polchinski, {\it Comm.\ Math.\ Phys.} {\bf 104}
                (1986) 
                37; O.\ Alvarez, {\it Nucl.\ Phys.} {\bf B216} (1983)
                125; 
                G.\ Moore, P.\ Nelson, {\it Nucl.\ Phys.} {\bf B266}
                (1986) 58.

\bibitem{ag}    P.\ Ginsparg, {Les Houches, Session XLIX, (1988)},
                Elsevier Science Publishers (1989). 

\bibitem{foeraur}  D.\ Foerster, {\it Nucl.\ Phys.} {\bf B283} (1987) 669;
                   E.\ Aurell, P.\ Salomonson, {\it Comm.\ Math.\ Phys.}
                   {\bf 165} (1994) 233 and hep-th/9405140.

\bibitem{pmppp}  P.\ Menotti, P.\ P.\ Peirano, {\it Phys. Lett.} {\bf
                 353B} (1995) 444; {\it Nucl.\ Phys.\ B (Proc.\ Suppl.) 
                 } {\bf 47} (1996) 633 and preprint 
                 IFUP--TH 4/96 (hep-th/9602002), to appear on Nucl.\
                 Phys.\ B. 

\bibitem{janke} C.\ Holm, W.\ Janke, preprint FUB-HEP-19-94
                (hep-lat/9501004) and  {\it Phys. Lett.} {\bf 375B}
                (1996) 69. 

\bibitem{holmj} W.\ Bock, J.C.\ Vink, {\it Nucl.\ Phys.} {\bf B438}
                (1995) 320; C.\ Holm, W.\ Janke, preprint FUB-HEP-17-95
                (hep-lat/9511029). 

\bibitem{jev}   A.\ Jevicki, M.\ Ninomiya,  {\it Phys.\ Rev.} {\bf
                D33} (1986) 1634.

\bibitem{gen}   P.\ Menotti, P.\ P.\ Peirano, preprint IFUP--TH 38/96
                (hep-th/9607071).  


\end{thebibliography}
\end{document}